\documentclass[fleqn,usenatbib]{mnras}

\usepackage{mathptmx}
\usepackage[T1]{fontenc}

\DeclareRobustCommand{\VAN}[3]{#2}
\let\VANthebibliography\thebibliography
\def\thebibliography{\DeclareRobustCommand{\VAN}[3]{##3}\VANthebibliography}

\usepackage{graphicx}	% Including figure files
\usepackage{amsmath}	% Advanced maths commands
\usepackage{amssymb}	% Extra maths symbols
\usepackage{txfonts}
\usepackage{hyperref}
\usepackage{natbib}
\usepackage[singlelinecheck=false,justification=justified]{caption}
\usepackage{array,boldline}
\usepackage{booktabs}
\usepackage{tabularx}
\usepackage[scientific-notation=true]{siunitx}
\usepackage{longtable}

\title[Do ETVs reliably predict circumbinaries for PCEBs]{Eclipse timing variations in  post-common envelope binaries: Are they a reliable indicator of circumbinary companions?}

\author[D. Pulley et al.]{
D. Pulley,$^{1}$\thanks{E-mail: david@davidpulley.co.uk}
I. D. Sharp,$^{1}$\thanks{E-mail: ian.sharp@astro-sharp.com}
J. Mallett$^{1}$\thanks{E-mail: john@astro.me.uk}
and S. von Harrach$^{1}$\thanks{E-mail: svharrach@gmail.com}
\\
$^{1}$British Astronomical Association, Burlington House, Piccadilly, London W1J 0DU, UK\\
}

\date{Accepted XXX. Received YYY; in original form ZZZ}

\pubyear{2022}

\begin{document}
\label{firstpage}
\pagerange{\pageref{firstpage}--\pageref{lastpage}}
\maketitle

\begin{abstract}
{Post-common envelope binary systems evolve when matter is transferred from the primary star at a rate that cannot be accommodated by its secondary companion.  A common envelope forms which is subsequently ejected resulting in a system with a binary period frequently between 2 and 3 hours. Where circumbinary companions are predicted, it remains unclear whether they form before or after the common envelope ejection. From observations of eclipse time variations (ETVs), exoplanet databases e.g. NASA Exoplanet Archive, list typically a dozen systems with confirmed circumbinary planets. Here we examine seven of these systems, discuss other possible causes and consider whether, for these dynamic systems, the ETV methodology is a reliable indicator of planetary companions. The systems selected were those where we could determine precise eclipse timings, free from significant extraneous effects such as pulsations, and present 163 new times of minima permitting us to test existing models. Over thirty circumbinary models have been proposed for these seven systems and note all, other than the latest model for NY Vir which remains to be fully tested, fail within a year to accurately predict eclipse times. In examining alternative mechanisms we find that magnetic effects could contribute significantly in two of the seven systems studied. We conclude that the structure of these dynamic systems, with the extreme temperature differences and small binary separations, are not fully understood and that many factors may contribute to the observed ETVs.}
\end{abstract}

\begin{keywords}
binaries: close --
             binaries: eclipsing --
             planetary systems --
             planets and satellites: formation --
             subdwarfs --
             white dwarfs
\end{keywords}

%%%%%%%%%%%%%%%%%%%%%%%%%%%%%%%%%%%%%%%%%%%%%%%%%%

%%%%%%%%%%%%%%%%% BODY OF PAPER %%%%%%%%%%%%%%%%%%

\section{Introduction}
Three decades have passed since the first confirmed exoplanet discovery \citep{wolszczan1992planetary} and today in excess of 5000 exoplanets have been confirmed and listed in the NASA Exoplanet Archive (NEA)\footnote{\ NASA\ Exoplanet\ Archive \url{https://exoplanetarchive.ipac.caltech.edu/}}.  Although many discovery methods have been developed to assist in the search for these new worlds, two methods account for over 95\% of these confirmations, transit (77\%) and radial velocity (18\%).  Contrary to expectations, the NEA and similar databases also show a strong bias towards exoplanets orbiting single star systems with fewer than 10\% being associated with multi-star systems. Of this small number the majority, possibly as many as 90\%, are in S-type orbits, orbiting just one of the stars in the system.

\setlength{\heavyrulewidth}{1.2pt}
\setlength{\abovetopsep}{4pt}

\begin{table*}
\caption{PCEBs with claimed circumbinary planets detected using the ETV methodology} 
\label{table_PCEBs}
\centering 
\begin{tabular}{l l l l l l l c}
\toprule
\multicolumn{4}{c}{} & \multicolumn{2}{c}{No. of Planets} &  & Reported in\\
PCEB System & Type & RA & Dec. & NEA & EPA & Reference Source & this paper? \\
\midrule 
2MASS J19383260+4603591 & PCEB/sdB & 19 38 32.61 & +46 03 59.1 & 3 & 3 & \cite{esmer2022detection} & No\\      DE CVn & PCEB/WD & 13 26 53.27 & +45 32 46.7 & 1 & N/L & \cite{han2018cvn} & Yes\\
HS0705+6700 & PCEB/sdB & 07 10 42.07 & +66 55 43.6 & N/L & 2 & \cite{sale2020eclipse} & Yes\\
KIC 10544976 & PCEB/WD & 19 42 37.19 & +47 45 48.4 & N/L & 1 & \cite{almeida2019orbital} & No\\
HW Vir & PCEB/sdB & 12 44 20.24 & -08 40 16.8 & 1 & 1 & \cite{beuermann2012bquest} & Yes\\
NN Ser & PCEB/WD & 15 52 56.13 & +12 54 44.4 & 2 & 2 & \cite{marsh2014planets} & Yes\\
NSVS 14256825 & PCEB/sdB & 20 20 00.46 & +04 37 56.5 & 1 & 1 & \cite{zhu2019close} & Yes\\
NY Vir & PCEB/sdB & 13 38 48.14 & -02 01 49.2 & 2 & 2 & \cite{song2019updated} & Yes\\
OGLE-GD-ECL-11388 & PCEB/sdB & 13 34 49.87 & -64 08 31.8 & N/L & 1 & \cite{hong2016photometric} & No\\
RR Cae & PCEB/WD & 04 21 05.56 & -48 39 07.1 & 1 & 1 & \cite{qian2012acircumbinary} & Yes\\
\bottomrule
\end{tabular}
\end{table*}

The small number of exoplanets orbiting multi-star systems is due in part to observational bias and is the subject of current research \citep{matson2018stellar, bonavita2020frequency, fontanive2021census}. Bonavita (2020) found no significant difference in exoplanet frequency between single and binary star systems but did conclude there was evidence to suggest fewer exoplanets were to be found in close binary systems.

Unsurprisingly the majority of exoplanets in multi-star systems have been discovered by either the transit or radial velocity methods contributing some 92\% of the total.  Direct imaging, microlensing, pulsar timing and transit timing variations make up half of the remaining discoveries whilst ETVs provide the remaining 4\% with 16 confirmed exoplanets listed on the NEA database. Other exoplanet databases such as the Extrasolar Planets Encyclopedia (EPE)\footnote{Extrasolar Planets Encyclopedia \url{http://exoplanet.eu/}} draw similar conclusions to those of the NEA although absolute numbers differ.

Whilst eclipse times of isolated non-contact eclipsing binary systems can be predicted with precision, if their centre of mass is perturbed for example by the presence of a circumbinary third body, eclipses will occur earlier or later than expected. These ETVs can be translated into observed minus calculated, O--C, diagrams and so provide a unique method for detecting long period exoplanets, a technique underpinned by precision timing of the primary eclipses, see for example \cite{lee2009sdb+, qian2013search}.

The NEA lists 12 multi-star systems having 16 exoplanets which have been discovered by ETVs, including seven post-common envelope binaries (PCEBs) with either white dwarf (DE CVn, NN Ser, RR Cae) or subdwarf (2M1938+4603, HW Vir, NSVS 14256825, NY Vir) primaries. In addition, and not listed on the NEA, are many claims for circumbinary planets orbiting other PCEBs and examples include a possible circumbinary brown dwarf orbiting HS2231+2441, \cite{qian2010orbital}, and more recently \cite{wolf2021possible} who suggested a possible circumbinary planet orbiting NSVS 07826147.  Also many of the confirmed PCEB exoplanets listed in the NEA and similar catalogues have undergone an evolving history with each prediction being one in a sequence of relatively short lived circumbinary models, for example NY Vir has seen at least five models published during the past decade.

Planetary evolution in these systems is not fully understood but it is thought that after the more massive primary component enters its red giant phase and fills its Roche lobe, matter is transferred from the primary at a rate which cannot be accommodated by its smaller main-sequence companion and so forms a common envelope. As a consequence angular momentum is transferred from the binary system to the surrounding envelope bringing the stars closer together and reducing their binary period to a few hours. Ultimately the primary star evolves into either a subdwarf (sdB) or white dwarf (WD) and the common envelope is ejected.  Whether circumbinary creation is first generation, where planets form before and survive the common envelope ejection, or second generation and form in the short period after the envelope ejection, remains an unresolved question \citep{bear2014first}.

Other processes have been put forward to explain the apparent periodicity in the O--C diagrams and include (i) apsidal motion (ii) magnetic effects (iii) angular momentum loss (AML) through gravitational radiation \citep{paczynski1967gravitational} or through magnetic braking \citep{rappaport1983new}. However, none of these alternatives have yet been able to satisfactorily explain the magnitude of the observed ETVs.  Also, when a system's underlying ephemeris contains a positive quadratic coefficient, a circumbinary body with a long period is often invoked as providing the only consistent explanation for this orbital period increasing term, see for example \citep{qian2013search}.

In this paper we evaluate seven PCEBs with claimed circumbinary planets detected using the ETV methodology and listed in either or both the NEA and EPE databases, see Table \ref{table_PCEBs}. We present our new observations in Section 2 and discuss the long term circumbinary trends shown by these systems in Section 3.  Possible causes of the O--C variations are explored in Section 4 and our conclusions presented in Section 5.

\section{Observations and data reduction}
We report 163 new times of minima of seven sdB and WD PCEB systems observed between 2018 January and 2022 March using the telescopes and filters listed in Appendices \ref{table_A1} and \ref{table_A2}. The effects of differing atmospheric extinctions were minimised by making all observations at altitudes greater than 40 degrees and where the Moon's effulgence would not cause a light gradient across the image frames. All images were calibrated using master dark, flat and bias frames and then analysed with either AstroArt\footnote{AstroArt, \url{http://www.msb-astroart.com/}} or MaxIm DL\footnote{MaxIm\ DL, \url{http://www.diffractionlimited.com/}} software packages. The source flux was determined with aperture photometry using a variable aperture, whereby the radius is scaled according to the FWHM of the target image. Variations in observing conditions were accounted for by determining the flux relative to comparison stars in the field of view. 

To maximise signal to noise ratio with the shortest possible exposure time, many observations were made with either no, or a broadband, filter.  Where filters were used the apparent magnitude of the target was derived from the apparent magnitudes of the comparison stars and the average magnitude of the target calculated by the software. The comparison stars' catalogue magnitudes for the various filters were taken from the American Association of Variable Star Observers (AAVSO) Photometric All Sky Survey (APASS) catalogue or, in some cases, the US Naval Observatory CCD Astrograph Catalog, UCAC4, and were chosen to be similar to the target magnitudes. As far as possible comparison stars with similar colour indices to the target stars were selected.

During the processing of the images, checks were made for potential causes of measurement error including star roundness to ensure effective guiding and the images were free from aircraft and satellite trails.  The signal to noise ratio of all measurements and the deviation of the reference stars from the reference magnitude was also checked for each measurement.  Finally, the check star reference magnitude was confirmed to be within a tolerance range of the reference magnitude.

All of our new timings used in this analysis were first converted to barycentric Julian date dynamical time (BJD\_TBD) using either the time utilities of the Ohio State University\footnote{Ohio State University, \url{http://astroutils.astronomy.ohio-state.edu/time/}} or from custom-written Python code using the Astropy\footnote{Astropy, \url{https://www.astropy.org/}} library. For most data we calculated the times of minima using the procedure of \cite{kwee1956method} coded in our own Python utilities.  For those systems with a WD primary and exhibiting a flat bottom to the eclipse curve, we derived the best fit equations to the ingress and egress with linear regression before computing the eclipse midpoint.

Our new timings were combined with previously published times of minima and, where appropriate, the historic times were converted to BJD\_TBD before computing new linear and/or quadratic ephemerides and O--C residuals.

\section{PCEB circumbinary models and new data}
Analysis of binary star systems ETVs has become a standard methodology for detecting circumbinary objects.  Proposed by \cite{schneider1990photometric}, many investigators have subsequently employed this technique to identify potential circumbinary companions, confirmed exoplanets being listed in one of many online exoplanet catalogues.  Table \ref{table_PCEBs} lists the confirmed PCEB exoplanets discovered using ETV methodology and listed in two such catalogues, the NEA and the EPE, illustrating a high level of commonality between the two databases.

In the following sections we discuss past and present models for seven of the ten systems listed in Table \ref{table_PCEBs}, including in each model our latest observations.  A brief overview of many of these systems can be found in \cite{zorotovic2013origin} and \cite{lohr2014period}.

\subsection{HW Vir (BD-07 3477)}
\noindent HW Vir is considered the prototype of the HW Vir-like systems consisting of an eclipsing sdB primary star with an M5/6 dwarf secondary exhibiting quasi-sinusoidal period changes. Since its discovery there have been in excess of six models put forward to explain the observed period changes with the inclusion of up to four circumbinary objects.  Some investigators consider these solutions dynamically unstable but currently one circumbinary planet, HW Vir b, is identified on both the NEA and EPE databases.

HW Vir's eclipsing nature was discovered by \cite{menzies1986} and \cite{kilkenny1994period} first noted changes in the system's orbital period.  Observations over the next decade,  \cite{ccakirli1999ubvr}; \cite{kilkenny2003sdb} and \cite{ibanoglu2004high}, suggested the possible presence of a circumbinary brown dwarf with a period of between 18.8 yr and 20.7 yr but did not dismiss other possible causes of the period change. \cite{qian2008magnetic} also suggested a brown dwarf with a period of 15.7 yr. with a possible long period second circumbinary.

\begin{figure*}
\centering
   \includegraphics[width=17cm]{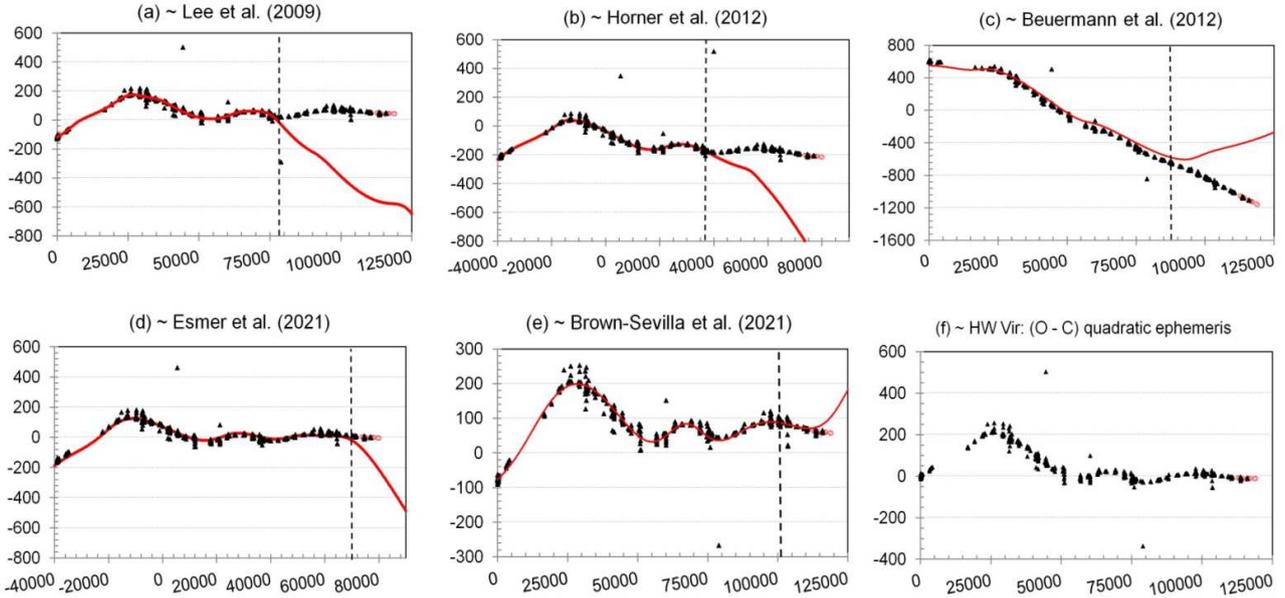}
     \caption{O--C plots for HW Vir.  Vertical axis is seconds and horizontal axis cycle number.  Historical data $\mathrm{\sim}$ black solid triangles; our new data $\mathrm{\sim}$ red open circles; circumbinary model $\mathrm{\sim}$ red line.  The original reported data is shown to the left of the vertical dotted line.  Note for Figs. \ref{HW_Vir_Charts}b and \ref{HW_Vir_Charts}d the reference epoch has been redefined by 38980 cycles ($\mathrm{\sim}$12.4 yr.).}
     \label{HW_Vir_Charts}
\end{figure*}

\cite{lee2009sdb+} were the first to propose that the ETVs were due to two massive bodies, 19.2 M${}_{J}$ and 8.5 M${}_{J}$, with orbital periods of 9.1 yr and 15.8 yr after ruling out other possible causes.  Lee's model, which includes a negative quadratic term, is shown in Fig. \ref{HW_Vir_Charts}a together with subsequent observations, where departure from this model is seen to occur around mid 2008.

\cite{horner2012dynamical} conducted a dynamical analysis of the Lee model and noted `... the system is so unstable that the planets proposed simply could not exist, due to a mean lifetime of less than a thousand years...'.  Their re-analysis proposed a revised two planet model, Fig. \ref{HW_Vir_Charts}b, with minimum masses and periods of 12 M${}_{J}$/4021 d and 11 M${}_{J}$/7992 d respectively.  Whilst this provided an improvement in orbital stability they noted that the system was still dynamically unstable and other mechanisms must be present.  It is noted that they chose to redefine the reference epoch introducing an offset of 38980 cycles, equivalent to 4549.73 days and that their new model fails almost immediately to predict future eclipse times.

A new circumbinary model was proposed by \cite{beuermann2012bquest} comprising of a linear ephemeris with two circumbinary objects (i) a planet of mass 14.3 M${}_{J}$ and a period of 12.7 yr. and (ii) a brown dwarf or low mass star of mass 30-120 M${}_{J}$ and a period fixed at 55 yr.  This model is recorded in the NEA database as a triple star system with one planet having a period of 12.7 yr. but recorded in the EPE database as a binary system with a single circumbinary planet.  As with the earlier models, the Beuermann model quickly failed as new observations were added, Fig. \ref{HW_Vir_Charts}c.

More recently the stability of circumbinary objects around HW Vir has also been investigated by \cite{esmer2021revisiting}, \cite{brown2021new} and \cite{mai2022eclipse}. Esmer included both new data from \cite{baran2018pulsations} and their data through to 2019 March.  They evaluated six possible models, all of which had stability issues, favouring a quadratic ephemeris with two circumbinary objects of minimum masses of 25 M${}_{J}$ and 13.9 M${}_{J\ }$and periods of 31.2 yr. and 13.6 yr. respectively.  However we find our new data post 2019 June does not fit predictions from their preferred model, Fig. \ref{HW_Vir_Charts}d.

Brown-Sevilla explored four models with two circumbinary companions, predominately brown dwarfs or low mass stellar objects, but found all to be dynamically unstable and were unable to confirm the \cite{beuermann2012bquest} stability claim of 10${}^{8\ }$yr.  We show, Fig. \ref{HW_Vir_Charts}e, their best fit for their PIKAIA+LM model with two brown dwarfs as companions where almost immediately new data fails to conform to predictions.  \cite{mai2022eclipse} found similar stability issues with their three and four circumbinary companion models with their three companion model being stable for at least 10 Myr but giving a poor fit and their four companion model having a good fit but highly unstable.  As a consequence they have not listed either model.

We record 18 new observations of the primary minima between 2019 June and 2022 February and find a quadratic ephemeris provides a marginally better fit to the data, Eq. \ref{HW_Vir_ephem_lin}
   \begin{equation}\label{HW_Vir_ephem_lin}
   \begin{aligned}
      BJD_{\mathrm{minQ}} ={} & 2445730.5567(3) + 0.116719542(6)E - \\
      & 1.85(36) \times 10^{-13}E^2
   \end{aligned}
   \end{equation}
\noindent where we have adopted the reference epoch of Lee (2009) and have included our quadratic residual O--C chart in Fig. \ref{HW_Vir_Charts}f.
 
\subsection{NY Vir (PG1336-018)}
NY Vir was identified as a pulsating short period eclipsing sdB/M5 binary system by \cite{kilkenny1998ec}.  Since its discovery there have been five published circumbinary models and a two planet model, NY Vir b and NY Vir c, is listed on both the EPE and NEA. 

\begin{figure*}
\centering
   \includegraphics[width=17cm]{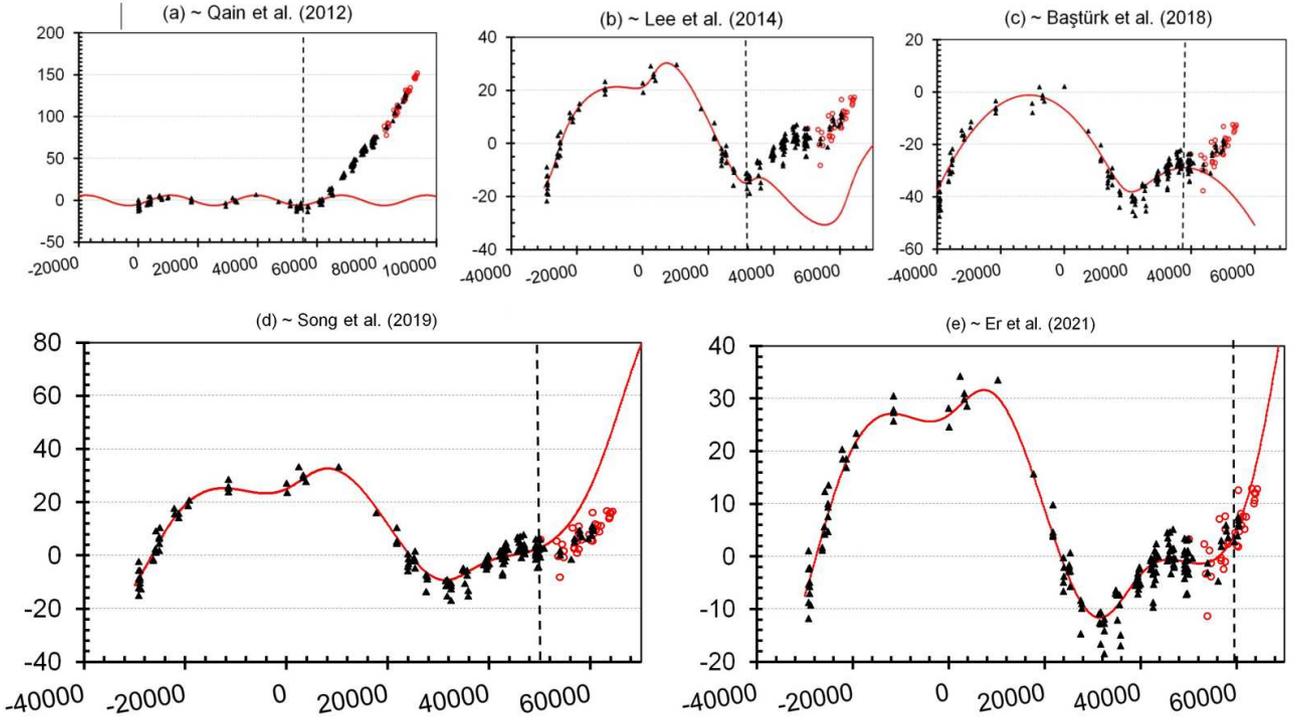}
     \caption{O--C plots for NY Vir.  Vertical axis is seconds and horizontal axis cycle number.  Historical data $\mathrm{\sim}$ black solid triangles; our new data $\mathrm{\sim}$ red open circles; circumbinary model $\mathrm{\sim}$ red line.  Note original reported data is shown to the left of the vertical dotted line.}
     \label{NY_Vir_Charts}
\end{figure*}

\cite{qian2012bcircumbinary} proposed the first circumbinary model having eliminated other possible mechanisms e.g. Applegate's magnetic quadrupole moment, AML, apsidal motion, indicating the presence of a giant planet with mass of 2.3 M${}_{J}$ and orbital period of 7.9 yr.  They noted a continuous period decrease, suggesting the presence of a possible second circumbinary object.  Their model included observations through to 2011 May and displayed a good fit to the data with a $\chiup$${}^{2}$ value of $\mathrm{\sim}$1.2 and an \textit{F}-test significance of above the 99.99\% level.  However, by early 2013 predictions made by this model failed to match future observations, Fig. \ref{NY_Vir_Charts}a.

A second model proposed by \cite{lee2014pulsating} considered both single and dual circumbinary systems.  Their single circumbinary model with a quadratic ephemeris differed from the earlier Qian model, allowing the planet eccentricity to be non-zero.  Although this model gave a low $\chiup$${}^{2}$${}_{red}$ of 1.31, no satisfactory explanation could be found for the quadratic term.  In contrast, their preferred two planet model had a $\chiup$${}^{2}$${}_{red}$ of 1.09, planetary masses of 2.8 M${}_{J}$ and 4.5 M${}_{J}$ and periods of 8.2 yr and 27.0 yr respectively. The periods were in a mean motion resonance of 3:10 and had orbital stability of $\mathrm{\sim}$850,000 yr.  Whilst the orbital stability was lower than expected it was recognised that their data only spanned two thirds of the second planet's orbital period.  This model fitted the known data but failed within twelve months of publication, Fig. \ref{NY_Vir_Charts}b.

After considering a number of possible solutions, \cite{bacsturk2018orbital} found that a single circumbinary planet with an underlying quadratic ephemeris gave the best fit to the observational data.  Their model yielded a $\chiup$${}^{2}$${}_{red}$ of 0.69 and was significantly different from the two previous models having a minimum planetary mass of 3.4 M${}_{J}$ but notably a longer orbital period of 20.63 years.  Their new data extended the observational baseline to 2017 June, providing almost one full orbit of the purported planet.  However, observations post 2019 July showed a significant departure from this hypothesis, Fig. \ref{NY_Vir_Charts}c.

A fourth model proposed by \cite{song2019updated} consisted of two circumbinary planets with an underlying quadratic ephemeris extending the baseline to 2018 April.  The masses and periods of the two planets were determined to be 2.66 M${}_{J}$ and 5.54 M${}_{J\ }$and 8.64 yr and 24.09 yr respectively and with similar orbital eccentricities of 0.15.  However, our new data indicates this model failing post 2019 August, Fig. \ref{NY_Vir_Charts}d.

The most recent two planet circumbinary model, \cite{er2021new}, bears similarities to that of \cite{song2019updated} with planets of mass 2.78 M${}_{J\ }$and 4.49 M${}_{J}$ and periods 8.18 yr and 27.0 yr. In this paper we report another 36 times of primary minima spanning 2018 January to 2022 March and extending the Er timeline from 2021 March.  Our recent data tends to fall below their new model, Fig. \ref{NY_Vir_Charts}e, but more observations over the coming months will be necessary to confirm or otherwise their predictions. Our revised quadratic ephemeris generates an O--C diagram very similar to that of \cite{er2021new} shown in Fig. \ref{NY_Vir_Charts}e.

\subsection{HS0705+6700 (V470 Cam)}
HS0705+6700 was identified as a 14.7 magnitude sdB star from the Hamburg Schmidt Quasar Survey and confirmed by \cite{drechsel2001hs} as a short period ($\mathrm{\sim}$2.3hr) eclipsing PCE sdB+dM binary system with the components separated by 0.81R${}_{o}$.  The mass of the sdB as derived from spectroscopy is 0.48M${}_{o}$ and light curve analysis suggests a mass ratio of 0.278 yielding the mass of the secondary as 0.13M${}_{o}$ and separation of 0.81R${}_{o}$.  For the well-defined binary period of 0.0956466d, orbital mechanics suggests a stellar separation of 0.75R${}_{o}$, not 0.81R${}_{0\ }$specified.\footnote{\ H.\ Drechsel,\ private\ communication with J. Mallett\ 2021\ May} 

Since its discovery, eight circumbinary models have been proposed but, within twelve months of publication, all models fell short in predicting this system's long-term behaviour.  Nonetheless one model, \cite{sale2020eclipse}, is listed on the EPE database.

\begin{figure*}
\centering
   \includegraphics[width=17cm]{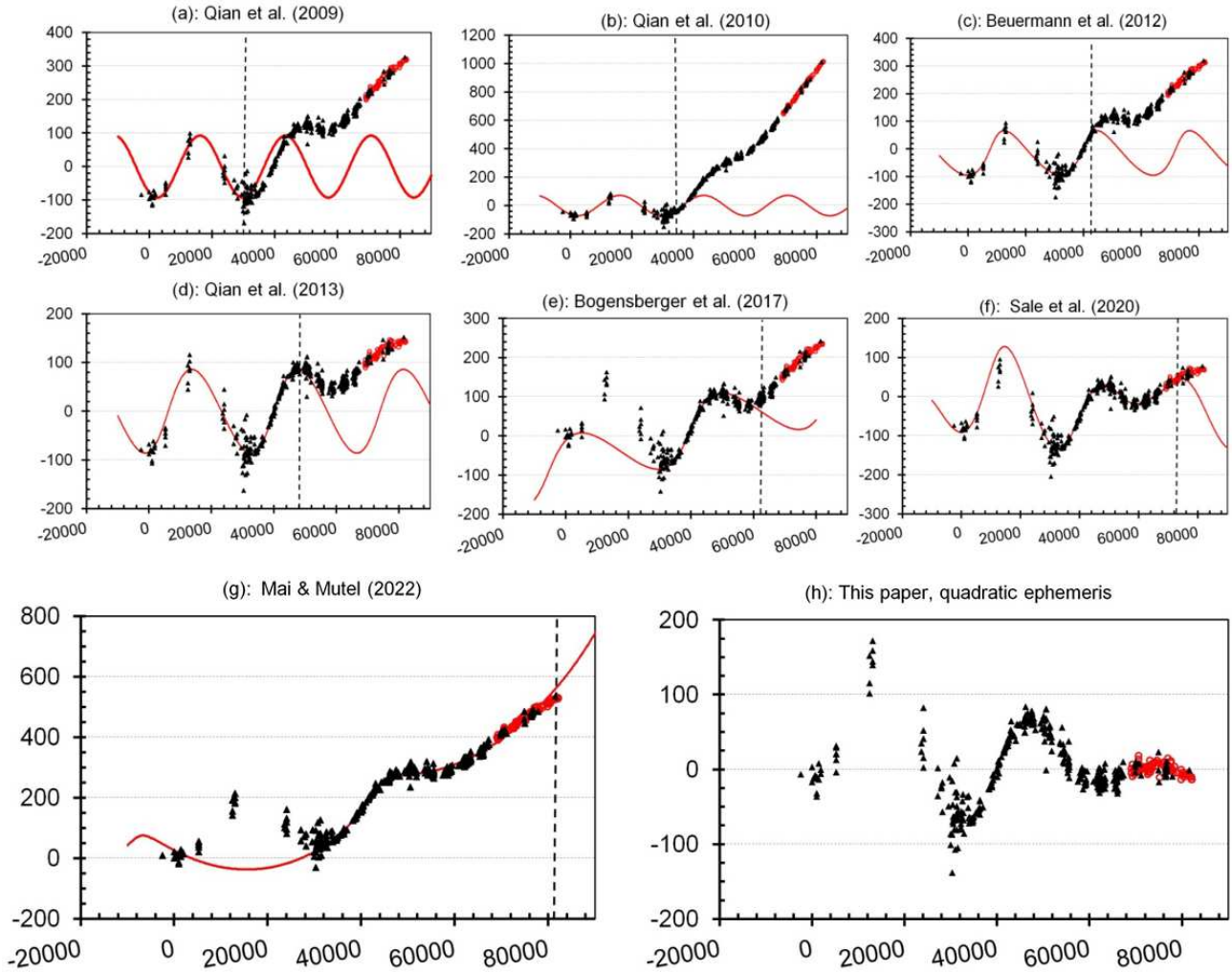}
     \caption{O--C plots for HS0705+6700.  Vertical axis is seconds and horizontal axis cycle number.  Historical data $\mathrm{\sim}$ black solid triangles; our new data $\mathrm{\sim}$ red open circles; circumbinary model ~$\mathrm{\sim}$ red line.  Note original reported data is shown to the left of the vertical dotted line.}
     \label{HS0705+6700_Charts}
\end{figure*}

With some eight years of data, \cite{qian2009detection}, predicted the presence of a circumbinary brown dwarf in a 7.15 yr circular orbit but, Fig. \ref{HS0705+6700_Charts}a, this model failed as new data became available.  A revised model was proposed by \cite{qian2010orbital} with a lower mass brown dwarf in a 7.15 yr circular orbit, inadvertently specified as 15.7 yr, but by late 2010 new data showed this model could no longer represent this system, Fig. \ref{HS0705+6700_Charts}b.

Whilst \cite{qian2009detection}) and (2010) proposed a circumbinary brown dwarf in a circular orbit,  \cite{ccamurdan2012photometric} and \cite{beuermann2012aquest} both proposed a linear ephemeris with a single circumbinary object in an elliptical orbit.  \c{C}amurdan does not provide either the orbital eccentricity nor the time of periastron for their model making it difficult to reconstruct their O--C plot.  Beuermann, however, predicts a circumbinary brown dwarf third body of mass 31.5/sin\textit{i} M${}_{J}$${}^{\ }$and orbital eccentricity of${}^{\ }$0.38.  Their O--C plot is shown in Fig. \ref{HS0705+6700_Charts}c and again, with new data, their model failed by mid 2012.

The next model to be proposed, \cite{qian2013search}, suggested a quadratic ephemeris with a single brown dwarf circumbinary object of mass 32/sin\textit{i} M${}_{J}$ and orbital eccentricity of 0.19.  The positive coefficient of the quadratic term, 4.9 x 10${}^{-13}$,  ruled out AML through either gravitational waves or magnetic breaking, leading to the speculation that this may be part of a second long period circumbinary substellar object.  However within 18 months this model also failed to make reliable predictions, Fig. \ref{HS0705+6700_Charts}d.  \cite{pulley2015eclipse} produced a similar model to that of Qian (2013) but noted in their Addendum that their new observations did not align with predictions.

Two recent models have been proposed by the University of Oxford research group, \cite{bogensberger2017further} and \cite{sale2020eclipse}.  Bogensberger proposed a linear ephemeris with a single circumbinary brown dwarf in an eccentric orbit of 0.37 with a period of 11.77 yr, substantially longer than suggested by earlier models.  However, new observations did not follow the model's predictions, Fig. \ref{HS0705+6700_Charts}e. Sale's 2020 model, and listed in the EPE database, differed in many respects from that of Bogensberger, including an ephemeris with a quadratic term and two circumbinary brown dwarfs in elliptical orbits with periods of 7.9 yr and 13.3 yr, Fig. \ref{HS0705+6700_Charts}f.  Again, with new observations, this model was unable to predict future eclipses beyond 2020 March.  Note, Bogensberger's times of minima are exposure start times; their amended mid-eclipse times are listed in Table A1 of \cite{sale2020eclipse}.

\cite{mai2022eclipse} explored a one and two companion model finding both were stable but with the two companion fit no better than that of the one companion model.  As a consequence they preferred the one component model, Fig. \ref{HS0705+6700_Charts}g, where it is seen that our new data departs from this model.

We report a further 47 times of minima observed between 2019 September and 2022 March, and find a positive coefficient quadratic ephemeris, Eq. \ref{HS0705_ephem_quad}, provides the best fit
   \begin{equation}\label{HS0705_ephem_quad}
   \begin{aligned}
      BJD_{\mathrm{minQ}} ={} & 2451822.7607(1) + 0.095646620(4)E + \\
      & 7.7(5) \times 10^{-13}E^2
   \end{aligned}
   \end{equation}
The O--C diagram is shown in Fig. \ref{HS0705+6700_Charts}h where the shape has the appearance of a damped periodic wave.

\subsection{NSVS 14256825}
NSVS 14256825 was identified as a variable star in the Northern Sky Variability Survey (NSVS), \cite{wozniak2004northern}, and subsequently as an eclipsing sdB/dM binary by \cite{wils2007nsvs}. \cite{kilkenny2012detection}, noted a period increase but, with limited data, was unable to assign a cause.  Four circumbinary models have been proposed for this system which is listed on both the NEA and EPE databases with one circumbinary planet, \cite{zhu2019close} and \cite{nasiroglu2017there} respectively.  For completeness, we include the NSVS and All Sky Automated Survey (ASAS) data in the following charts, although most investigators omit these observations from their analyses.

\begin{figure*}
\centering
   \includegraphics[width=17cm]{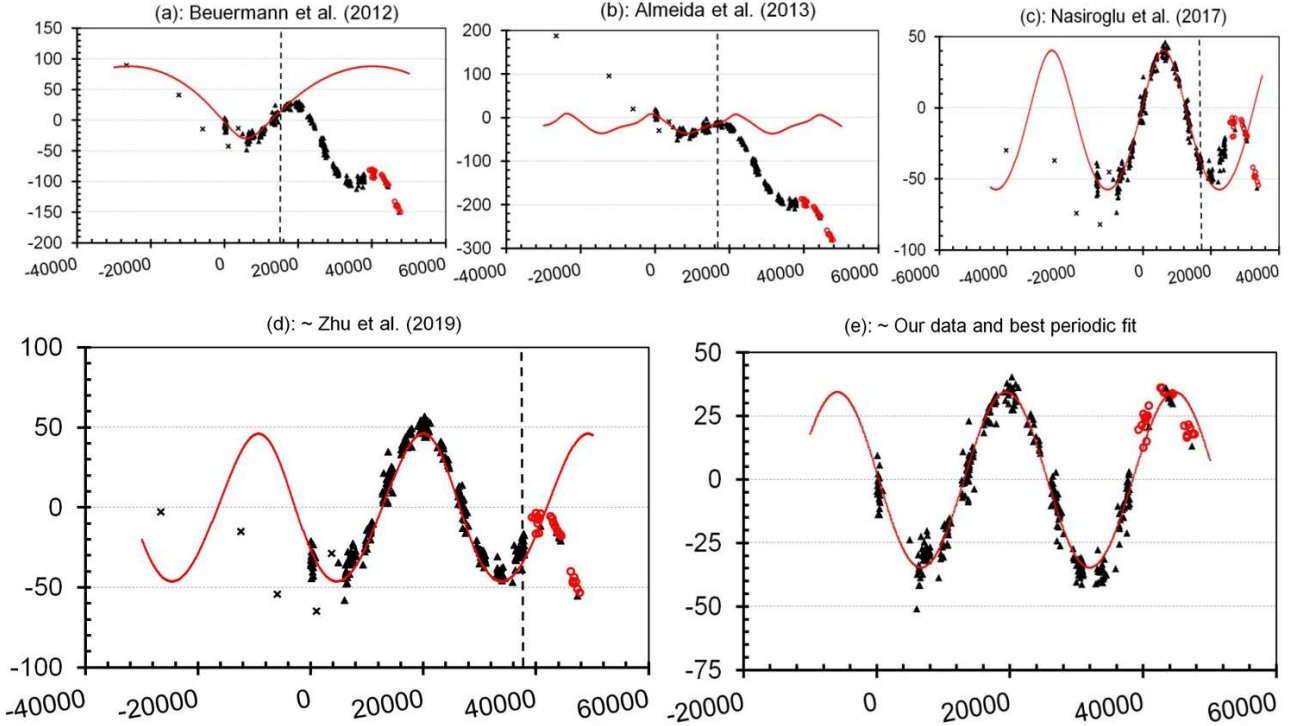}
     \caption{O--C plots for NSVS 14256825. Vertical axis is seconds and horizontal axis cycle number. Historical data ~$\mathrm{\sim}$ black solid triangles; our new data ~$\mathrm{\sim}$ red open circles; circumbinary model ~$\mathrm{\sim}$ red line. Black cross ~$\mathrm{\sim}$ NSVS and ASAS data. Note original reported data is shown to the left of the vertical dotted line..}
     \label{NSVS_14256825_Charts}
\end{figure*}

\cite{beuermann2012aquest} were the first to propose a single circumbinary planet of mass 12 M${}_{J}$ and a period of 20 yr in a 0.5 eccentric orbit.  Adopting a linear ephemeris, and including both NSVS and ASAS data in their analysis, they recognised their model was speculative, having little more than one half a period of circumbinary data. As can be seen from Fig. \ref{NSVS_14256825_Charts}a, their proposal cannot predict eclipse times more than a year ahead.

The following year \cite{almeida2013two}, proposed a model adopting a linear ephemeris with two circumbinary planets of mass 2.9 M${}_{J}$ and 8.1 M${}_{J}$, periods of 3.5 yr and 6.9 yr and orbital eccentricities of 0.0 and 0.52 respectively.  Almost immediately this model failed to predict future eclipses, Fig. \ref{NSVS_14256825_Charts}b. Subsequent investigators noted their two planet solution was extremely unstable on time-scales of less than a thousand years, \cite{wittenmyer2013dynamical}, and had insufficient timing data to constrain the best-fit parameters, \cite{hinse2014revisiting}.

Excluding NSVS, ASAS and WASP (Wide Angle Search for Planets) datasets, \cite{nasiroglu2017there} proposed a single circumbinary brown dwarf of mass greater than 15 M${}_{J}$ in a 9.9 yr elliptical orbit with eccentricity of 0.175, having first considered and eliminated the \cite{applegate1992mechanism} effect.  Their model utilised a linear ephemeris but with the reference epoch offset by 13768 cycles from that used by other investigators, where we see that this model also departs from new observations taken over the following 18 months, Fig. \ref{NSVS_14256825_Charts}c.

A further single circumbinary model was proposed by \cite{zhu2019close} with a circumbinary planet/brown dwarf of minimum mass of 14.2 M${}_{J}$, a period of 8.83 yr and an orbital eccentricity of 0.12.  Their model provided a better fit to the observations than that of \cite{nasiroglu2017there}, Fig. \ref{NSVS_14256825_Charts}d, but with our new data this model failed within 12 months of publication.  More recently \cite{wolf2021possible} were unable to fit a single circumbinary object to the recorded data, suggesting at least two circumbinary bodies and additional observations were needed to explain the O--C variations.

We provide 20 new times of primary minima and together with all existing data we determined both a linear and quadratic ephemeris.  Because of the poor precision of both the NSVS and ASAS data, typically 45s, we omitted these observations from our analysis.  We found the quadratic ephemeris provided the best fit to the data and noted its apparent sinusoid shape. Fitting a best sinusoid gave a $\chiup$${}^{2}$ of 5.3, Fig. \ref{NSVS_14256825_Charts}e, with the underlying ephemeris, Eq. \ref{NSVS_14256825_quad}
   \begin{equation}\label{NSVS_14256825_quad}
   \begin{aligned}
      BJD_{\mathrm{minQ}} ={} & 2454274.20875(5) + 0.110374157(4)E - \\
      & 1.26(9) \times 10^{-12}E^2
   \end{aligned}
   \end{equation}
\noindent and with sinusoidal parameters (i) amplitude 34.56s (ii) eccentricity 0.02 (iii) period 7.65 yr (iv) periastron angle 2.237 rads and (v) date of periastron 2456703.  Whilst the fit looks reasonable up to E $\mathrm{\sim}$40000, our new data post 40000 does indicate a departure from this prediction.   

 \captionsetup{width=17cm,justification=raggedright}

\subsection{NN Ser}
NN Ser was first identified as a pre-cataclysmic variable in 1989 and subsequently confirmed as a PCEB system comprising of a white dwarf primary with a very low mass, fully convective, M dwarf secondary companion.  With limited data \cite{brinkworth2006detection} identified a period decrease, assigning it to either AML through magnetic breaking or the presence of a third body.  Five circumbinary models have been published to explain the observed period change, with different models being included in the NEA and EPE databases.

\begin{figure*}
\centering
   \includegraphics[width=17cm]{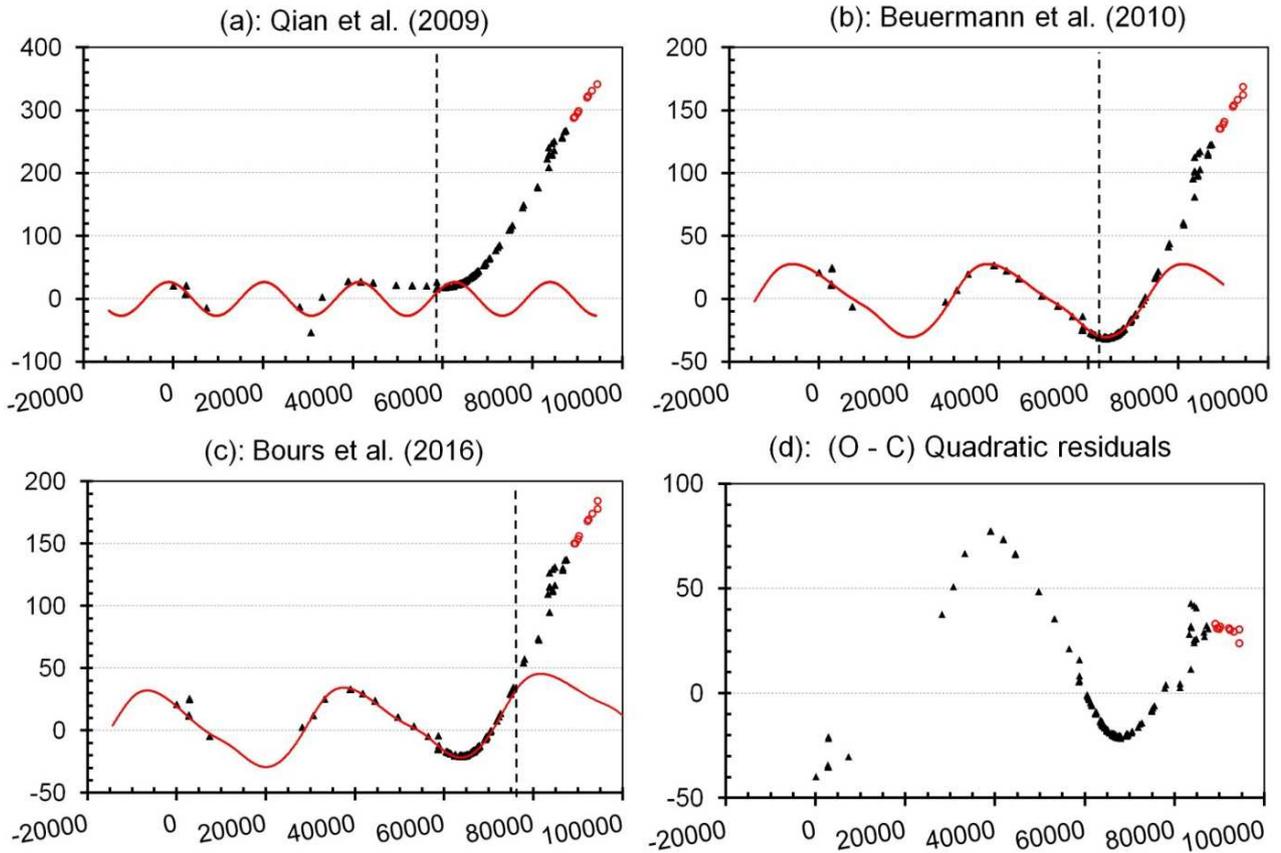}
     \caption{O--C diagrams for NN Ser.  Vertical axis is seconds and horizontal axis cycle number.  Historical data $\mathrm{\sim}$ black solid triangles; our new data $\mathrm{\sim}$ red open circles; circumbinary model $\mathrm{\sim}$ red line.  Note original reported data is shown to the left of the vertical dotted line.  Qian, Fig (a), was unaware of the data points around E $\mathrm{\sim}$55000.  The models transitioning from 2010 to 2016 via Beuermann (2013) and Marsh (2014) differ little from Beuermann et al. (2010) and Bours et al. (2016)}
     \label{NN_Ser_Charts}
\end{figure*}

\cite{qian2009detection} noted the presence of a cyclical change in the binary period and proposed a model consisting of a substellar companion of minimum mass 11.1 M${}_{J}$ in a circular orbit having a period of 7.56 yr, Fig. \ref{NN_Ser_Charts}a.  However from data not available to Qian, Parsons et al. (2010) indicated `the proposed third body does not exist'.

The possibility of multiple circumbinary bodies was first explored by \cite{beuermann2010two} with two planets in 2:1 resonance providing a marginally best $\chiup$${}^{2}$ fit.  They explored a single planet solution but noted other factors, e.g. the Applegate mechanism, must be present to explain the observed variations.  Beuermann's preferred solution identified two planets with minimum masses 2.3 M${}_{J}$ and 6.9 M${}_{J}$, periods 7.75 yr and 15.5 yr and orbital eccentricities of 0.2 and 0 respectively. 

Modifications to the \cite{beuermann2010two} planetary model were put forward first by \cite{beuermann2013quest} and then \cite{marsh2014planets}, the most significant addition being an elliptical orbit for the outer planet with minor changes to the remaining parameters.  These three models produced very similar residual plots and in Fig. \ref{NN_Ser_Charts}b we show the results published by \cite{beuermann2010two}.  It is of note that this model is recorded in the EPE database whereas the NEA database adopts the \cite{marsh2014planets} parameters.

The durability of Beuermann's 2010 model led to increasing confidence in the presence of two circumbinary planets, with NN Ser being one of a very few PCEBs with apparently stable planetary systems.  However observations reported by \cite{bours2016long} indicated that the underlying ephemeris needed to include a quadratic term, noting that  `The best test... will come in 2018-2019 when the model predicts a maximum and subsequent downturn in the O--C eclipse times'.  As seen from Fig. \ref{NN_Ser_Charts}c, Bours' 2018/19 downturn test failed, indicating that none of the proposed NN Ser models to date have been able to predict the long-term future of this system.

 We report a further 9 observations between 2020 March and 2022 February and compute new linear and quadratic ephemerides, with the quadratic ephemeris giving a significantly lower $\chiup$${}^{2}$,  Eq. \ref{NN_Ser_quad}.
   \begin{equation}\label{NN_Ser_quad}
   \begin{aligned}
      BJD_{\mathrm{minQ}} ={} & 2447344.52512(12) + 0.130080079(4)E  + \\
      & 7.7(4) \times 10^{-13}E^2
   \end{aligned}
   \end{equation}
The reduced $\chiup$${}^{2}$ is in excess of 250 indicating historical significant changes in binary period, Fig. \ref{NN_Ser_Charts}d.

\subsection{RR Cae}
RR Cae is an eclipsing binary comprising a white dwarf primary and a dM secondary with a 7.29 hr period and listed with one circumbinary planet on both the NEA and EPE databases.  First identified by \cite{krzeminski1984lft} as an eclipsing binary, further observations recorded by \cite{bruch1998eclipsing} suggested the orbit may be eccentric.  However \cite{maxted2007mass} did not support this, classifying the secondary as an M4 star. \cite{parsons2010orbital} recorded two more primary eclipse times, noting that any O--C variations could be caused by an Applegate type mechanism and making no claims for the presence of a circumbinary object.

\begin{figure*}
\centering
   \includegraphics[width=17cm]{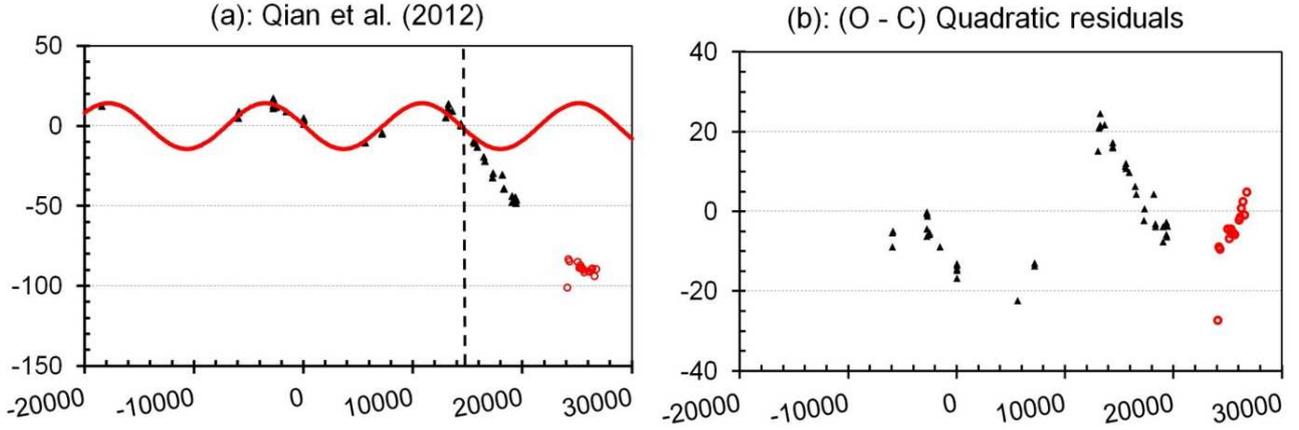}
     \caption{O--C plots for RR Cae.  Vertical axis is seconds and horizontal axis cycle number.  Historical data $\mathrm{\sim}$ black solid triangles; our new data $\mathrm{\sim}$ red open circles; circumbinary model $\mathrm{\sim}$~ red line.  Note original reported data is shown to the left of the vertical dotted line.}
     \label{RR_Cae_Charts}
\end{figure*}

\cite{qian2012acircumbinary} proposed a circumbinary model, underpinned with a positive coefficient quadratic ephemeris, to explain the observed cyclical O--C residuals, dismissing both apsidal motion through strong tidal interactions and the fully convective M4 secondary not supporting Applegate type mechanisms. Their circumbinary model comprised a single giant planet with minimum mass 4.2 M${}_{J\ }$, a period of 11.9 yr and LTT of 14.3s.  Since the positive quadratic coefficient could not be explained by AML they proposed the presence of a circumbinary companion with a long period.  Their model, listed on the EPE and NEA databases, is shown in Fig. \ref{RR_Cae_Charts}a but within 12 months new observations could not be explained by this model.

We report a further 16 times of minima recorded between 2019 December and 2022 February and compute both a linear and quadratic ephemeris. The quadratic ephemeris provides a better fit with a marginally lower $\chiup$${}^{2}$, Eq. \ref{RR_Cae_quad}.
   \begin{equation}\label{RR_Cae_quad}
   \begin{aligned}
      BJD_{\mathrm{minQ}} ={} & 2451523.04871(3) + 0.303703652(3)E - \\
      & 4.6(9) \times 10^{-13}E^2
   \end{aligned}
   \end{equation}
The O--C residuals computed from Eq. \ref{RR_Cae_quad} are shown in Fig. \ref{RR_Cae_Charts}b having a quasi-cyclical form   

\subsection{DE CVn}
DE CVn was identified as an eclipsing binary comprising a white dwarf primary and a late type main sequence star with a giant circumbinary planet listed on both the NEA and EPE databases, see \cite{han2018cvn}.  

First identified as an eclipsing binary by \cite{robb1997}, further observations from \cite{holmes2001}, \cite{tas2004times}, \cite{van2006cvn} and \cite{parsons2010orbital} gave no indication of the presence of any circumbinary object with \cite{lohr2014period} stating ``... the whole O--C diagram is fully consistent with a constant period...''.

\begin{figure*}
\centering
   \includegraphics[width=17cm]{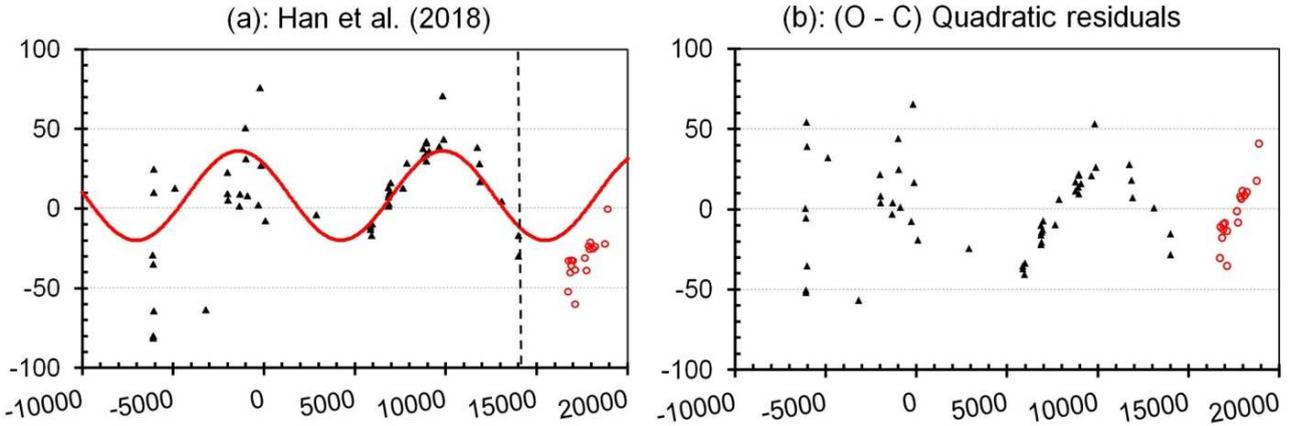}
     \caption{O--C plots for DE CVn.  Vertical axis is seconds and horizontal axis cycle number.  Historical data $\mathrm{\sim}$ black solid triangles; our new data $\mathrm{\sim}$ red open circles; circumbinary model $\mathrm{\sim}$ red line.  Note original reported data is shown to the left of the vertical dotted line.}
     \label{DE_CVn_Charts}
\end{figure*}

So far only \cite{han2018cvn} has derived a planetary model for this system, adopting a negative quadratic ephemeris to describe the underlying period change.  Their model indicates the presence of a third body of minimum mass of 10.5 M${}_{J}$ and 11.22 yr period producing an LTTE of 21.6s.  Their O--C representation is shown in Fig. \ref{DE_CVn_Charts}a where their analysis did not include data pre-2002 January, E $\mathrm{<}$ -1500, owing to the large uncertainties exceeding 60s.  We have included all earlier data in Fig. \ref{DE_CVn_Charts}a together with our new data, post 2020 January, where it can be seen these new observations do not fit the Han model.

We have included 17 new times of minima observed between 2020 January and 2022 March,  computed new quadratic ephemeris, Eq. \ref{DE_CVn_quad}, which included data pre-2002 January.
   \begin{equation}\label{DE_CVn_quad}
   \begin{aligned}
      BJD_{\mathrm{minQ}} ={} & 2452784.55410(5) + 0.364139445(10)E - \\
      & 1.74(6) \times 10^{-11}E^2
   \end{aligned}
   \end{equation}
The quadratic ephemeris provides a substantially lower $\chiup$${}^{2}$ of 26 opposed to 89 for the linear ephemeris, demonstrating a strong underlying quadratic distribution.  The O--C residuals for the quadratic ephemeris are shown in Fig. \ref{DE_CVn_Charts}b.  Whilst the early data does show expected scatter, a cyclical pattern is apparent in the data of the quadratic residuals.
\\

\section{ETVs and their origins}
We have observed seven PCE eclipsing binary systems which are listed on the NEA and/or EPE databases as having circumbinary companions and each having in excess of twenty years supporting eclipse timing data.  During the past fifteen years more than thirty circumbinary models have been proposed for these seven systems.  With one exception, NY Vir whose recent model still has to be tested, all models have proved incapable of predicting eclipse times more than a few years into the future.

Several processes put forward to explain the observed ETVs are considered below, including apsidal motion, AML through either gravitational radiation or magnetic breaking, Applegate type magnetic mechanisms and the presence of circumbinary companions.  The first three processes are frequently ruled out on grounds of their minimal contribution to the `cyclical' component of the observed ETVs, leaving circumbinary companions as the default mechanism.

\subsection{Apsidal motion}
Although apsidal motion can provide large cyclical ETVs, it has become the norm to dismiss this effect as the source of observed quasi-cyclical O--C variations seen in these PCE systems, e.g. \cite{lee2014pulsating}, \cite{sale2020eclipse}, \cite{mai2022eclipse}.  The close proximity of the binary components, typically less than a solar radius, should enable the binary orbits to circularize within a short timeframe although this could exceed the theoretical lifetime of the PCE system.  This appears to be confirmed with observations of secondary eclipses which, after making corrections for the R{\o}mer delay, occur close to phase 0.5, indicating orbital eccentricities of less than 0.001, for example see \cite{parsons2013timing}.  Small orbital eccentricities affect the timing of the eclipses, see \cite{barlow2012romer}, by approximately
\begin{equation} \label{barlow_eq} 
\Delta t\approx \ \frac{2}{\pi }{P_{bin}}ecos( \omegaup )
\end{equation} 
\noindent where $P_{bin}$ is the binary period, e is the orbital eccentricity and $\omegaup$ is the argument of periapsis.  For a typical PCEB with an orbital period of 0.1 d and an assumed eccentricity of 0.001, a maximum ETV of 5.5s could be observed.  However, as shown by \cite{parsons2013timing} for NN Ser, eccentricities may well be much less than 0.001.  In addition apsidal motion would provide a pure sinusoidal variation in ETVs, something that is not observed in these systems.  Whilst the evidence would suggest that apsidal motion is not the underlying cause of the observed ETVs in PCEBs, it is possibly a minor contributing factor.

\subsection{Angular momentum loss (AML)}
Angular momentum loss within PCE sdB and WD binaries is commonly attributed to either gravitational radiation or magnetic breaking.  With AML, the binary separation becomes smaller over time accompanied by a reduction in the binary period.  This in turn is reflected in the system ephemeris with the addition of a negative coefficient quadratic term.  The gradual reduction in binary period predicted by AML has no quasi-cyclical component and, as such, cannot explain the overall ETV cyclical behaviour.

\subsection{Magnetic mechanisms}
A mechanism to potentially explain the eclipse time variations seen in O--C diagrams was proposed by \cite{applegate1992mechanism}.  This mechanism relies on the fact that the secondary star is magnetically active and undergoes solar-like magnetic cycles which are strong enough to redistribute the angular momentum within the star. This leads to cyclical shape changes which affect the gravitational quadrupole moment. Since the orbit of the stars is gravitationally coupled to variations in their shape, this effect could explain the small modulations that are observed in the binary period.

Since the original work, other researchers, e.g \cite{lanza1998orbital}, \cite{brinkworth2006detection}, have refined and improved the Applegate model. \cite{volschow2016eclipsing} produced the ``two-zone'' model which showed that for a number of close binaries the energy, E${}_{A}$, required to generate the Applegate mechanism is far higher than the energy available, assumed to be the energy E${}_{2}$, radiated by the secondary in the time of the period modulation, P${}_{mod}$.
   \begin{equation}\label{applegate_equation_1}
	E_2=L_2\ P_{mod}\
   \end{equation}
\noindent where L${}_{2}$ is the luminosity of the secondary.

For the systems studied in this paper, the results are summarized in Table \ref{table_applegate}. In all cases the ratio E${}_{A}$/ E${}_{2}$ $\mathrm{>}$$\mathrm{>}$ 1, so the Applegate mechanism is very unlikely to explain the period variation.

In a recent paper \cite{lanza2020internal} presented a different magnetic mechanism that required much less energy. He proposed that a non-axiosymmetric gravitational quadrupole moment is created by an internal (non-axiosymmetric) magnetic field in the secondary. This leads to an exchange of angular momentum between the spin and orbital motion of the secondary. If the two stars are not perfectly synchronized this results in a periodic modulation of the orbital period of the binary system. The resulting change of rotational energy, $\Delta$E${}_{rot}$, following \cite{mai2022eclipse}, is
   \begin{equation}\label{applegate_equation_2}
	\mathrm{\Delta }E_{rot}=\ \frac{m\ a^2\ {\mathrm{\Omega }}^2\ \mathrm{\Delta }\mathrm{P}}{{3P}_{bin}}\
   \end{equation}
where m is the reduced mass of the binary, a is the radius of the binary orbit, $\Omega$ is the angular velocity of the secondary, approximately equal to 2$\piup$/ P${}_{bin}$, $\Delta$P is the period variation and P${}_{bin}$ is the binary period.

In the presence of a circumbinary object with a period P${}_{mod}$, the O--C diagram would show a semi-amplitude, K, related to the binary period variation \citep{volschow2016eclipsing}, as
   \begin{equation}\label{applegate_equation_3}
	\frac{\mathrm{\Delta }\mathrm{P}}{P_{bin}}\mathrm{=}\frac{\mathrm{4}\mathrm{\piup }\mathrm{K}}{P_{mod}}\
   \end{equation}

V\"{o}lschow et al. list the values of K and $P_{mod}$ for a number of close binaries, based on O--C data for the supposed circumbinary of the strongest influence. Table \ref{table_applegate} shows the calculated values for the systems in this study and the ratio $\Delta$E${}_{rot}$ / E${}_{2\ }$from equations \ref{applegate_equation_1} to \ref{applegate_equation_3}. The ratio of the Lanza energy $\Delta$E${}_{rot}$ required to the energy E${}_{2\ }$available is seen to be much larger than 1, except in the case of RR Cae and DE CVn. Hence, there is insufficient energy available, and so the Lanza mechanism is unlikely to explain the observed period variations, except in the above two cases. For RR Cae and DE CVn a part of the period variation could be due to the Lanza effect. 

\begin{table*}
\caption{Applegate energy ratios and parameters for the systems in this study, based on values in \protect\cite{volschow2016eclipsing}. DE CVn values are from \protect\cite{van2006cvn} and \protect\cite{han2018cvn}. Lanza energy ratios $\Delta$E${}_{rot}$ / E${}_{2\ }$calculated from equations \ref{applegate_equation_1} to \ref{applegate_equation_3}. $M_{\sun}$, $R_{\sun}$ and $L_{\sun}$ are the solar mass, radius and luminosity respectively.}
\newcolumntype{L}{>{\centering\arraybackslash}m{2.10cm}}
\newcolumntype{A}{>{\centering\arraybackslash}m{1.1cm}}
\newcolumntype{R}{>{\centering\arraybackslash}m{0.70cm}}
\newcolumntype{E}{>{\centering\arraybackslash}m{1.2cm}}
\newcolumntype{P}{>{\centering\arraybackslash}m{1.0cm}}
\label{table_applegate}
\centering 
\begin{tabularx}{\textwidth}{ 
   L | A | R | R | R | R | E | P | P | P | E | E}
\hlineB{2.0}
Binary system & Applegate ratio \newline E\textsubscript{A}/ E\textsubscript{2} & m/$M_{\sun}$ & a/$R_{\sun}$ & K (s) & P\textsubscript{mod} (yr) & L\textsubscript{2}/$L_{\sun}$ & P\textsubscript{bin} (d) & $\Delta$P/P\textsubscript{bin} (x10\textsuperscript{-7}) & E\textsubscript{2} /J (x10\textsuperscript{32}) & $\Delta$E\textsubscript{rot} /J (x10\textsuperscript{33}) & Lanza $\Delta$E\textsubscript{rot}/E\textsubscript{2}\\
\hlineB{2.0}\vspace{4pt}
HS0705+6700 & 140 & 0.105 & 0.810 & 67.0 & 8.41 & 0.0022 & 0.096 & 31.7 & 2.27 & 40.4 & 178.4\\
HW Vir & 108 & 0.110 & 0.860 & 563.0 & 55.0 & 0.0030 & 0.117 & 40.8 & 20.3 & 41.1 & 20.3\\
NN Ser & 64 & 0.092 & 0.934 & 27.65 & 15.5 & 0.0014 & 0.130 & 7.11 & 2.80 & 5.61 & 20.0\\
NY Vir & 106 & 0.096 & 0.770 & 12.20 & 7.90 & 0.0014 & 0.101 & 6.15 & 1.38 & 6.92 & 50.1\\
NSVS 14256825 & 101 & 0.083 & 0.740 & 20.00 & 6.86 & 0.0010 & 0.110 & 11.6 & 1.85 & 8.96 & 48.5\\
RR Cae & 59 & 0.129 & 1.620 & 7.20 & 11.90 & 0.0036 & 0.304 & 2.41 & 5.30 & 1.53 & 2.9\\
DE CVn & 26 & 0.227 & 2.090 & 14.00 & 11.22 & 0.0165 & 0.364 & 4.97 & 22.8 & 6.14 & 2.7\\
\hlineB{2.0}
\end{tabularx}
\end{table*}

\subsection{Circumbinary companions}
An often adopted philosophy when considering circumbinary modelling has been \lq...when you have eliminated the impossible, whatever remains, \textit{however improbable}, must be the truth.\rq\footnote{\ Sherlock\ Holmes\ in\ conversation\ with\ Dr.\ Watson,\ Sign\ of\ the\ Four,\ Sir\ Arthur\ Conan\ Doyle\ 1890}  For the systems considered here, and methodically eliminating apsidal motion, AML and Applegate-type mechanisms, we are left with circumbinary companions as the only possible explanation of the observed cyclical ETVs. However, this approach makes two assumptions (i) that all potential causes of ETVs are known and (ii) all known causes of ETVs are fully understood.

The PCEBs considered here question this approach with in excess of thirty models proposed for these seven systems, made over the past two decades, whilst all but one are incapable of predicting eclipse times more than a year into the future. The one exception is the recent, and still to be tested, new model for NY Vir, \cite{er2021new}.  Whilst this model was computed with data up until 2021 March our new data does appear to fall consistently below the model's forward predictions by some 5s.  However this small difference could be attributed to observational uncertainty and so it remains far too early to form any conclusion with further observations over the coming twelve months necessary to determine the validity of their model.

As new data becomes available a system's observational timeline is extended, in many instances this is now in excess of three decades, and the O--C plot becomes more complex and models with an increasing number of potential circumbinary companions are necessary to provide a good fit.  In principle it is possible to improve the fit by adding in more and more companions.  Early circumbinary models have relied upon one or possibly two companions together with a possible quadratic ephemeris.  However new models are now considering three and four circumbinary companions, some with orbital periods of hundreds of years.  Whilst these models may provide an improved fit to the O--C curve, they have so far exhibited poor orbital stability indicating that they do not provide a good long-term representation  of the observed ETVs.

For three systems, NY Vir, HS0705+6700 and NN Ser, the best fit to the O--C residuals is achieved by the inclusion of a quadratic term with a positive coefficient.  The presence of this term is difficult to explain through known mechanisms of AML, apsidal motion or magnetic effects leaving the inclusion of an additional long period circumbinary body as the only solution, see for example \cite{qian2013search}.  However, while this has been put forward as a working hypothesis it has still to be validated and, if two orbital cycles are needed to have confidence in a model, it may be many decades before this can be confirmed.

We note that a number of these systems, particularly HW Vir, HS0705+6700, and possibly NN Ser, Figs. \ref{HW_Vir_Charts}f, \ref{HS0705+6700_Charts}h and \ref{NN_Ser_Charts}d respectively, show a continuing decaying LTT amplitude in their O--C plots.  Whether this is an indicator of a change in behaviour is difficult to assess at this time but a further two or three decades of observations may bring greater clarity to these interesting stellar objects. 

Two systems, NSVS14256825 and DE CVn, indicate a cyclical O--C plot over some 1.5 cycles although the early data for DE CVn does show much scatter which is related to the large imprecision of these observations, typically greater than 100s.  Time averaging or weighting this early data may assist in a future analysis but until more data is available the assignment of one or two circumbinary companions may be premature.

\begin{figure}
	\captionsetup{width=8.50cm}
   \includegraphics[width=8.50cm]{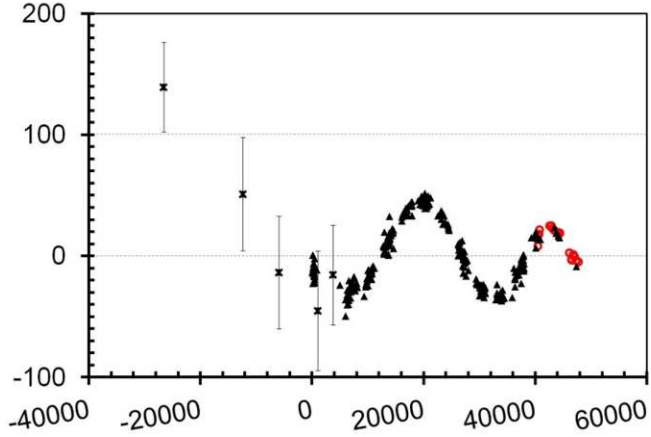}
     \caption{O--C chart for NSVS14256825 with both NSVS and ASAS data included in the analysis. Vertical axis is seconds and horizontal axis cycle number. NSVS \& ASAS data $\mathrm{\sim}$ black cross with error bars; other historical data $\mathrm{\sim}$ black solid triangles; our new data $\mathrm{\sim}$ red open circles.}
     \label{NSVS_14256825_all_data}
\end{figure}

Data that is considered to be of poor precision is often removed from a circumbinary analysis.  Whilst this data lacks precision, it may still be accurate and its omission can have a significant impact upon the analysis.  By way of an example \cite{beuermann2012aquest} included data sets derived from NSVS and ASAS observations in their analysis of NSVS14256825 extending the time line backwards 8 years, increasing the the observational period by some 40\%. Because of the lower precision of these datasets, most subsequent investigators chose to omit this data from their analysis resulting in a significantly different outcome.  In Fig. \ref{NSVS_14256825_Charts}e and Eq. \ref{NSVS_14256825_quad} we show the results of an analysis that omits both NSVS and ASAS data.  If we now include this data into our analysis we find a new ephemeris, Eq. \ref{NSVS_14256825_all_data_quad},

   \begin{equation}\label{NSVS_14256825_all_data_quad}
   \begin{aligned}
      BJD_{\mathrm{minQ}} ={} & 2454274.20895(4) + 0.110374134(4)E - \\
      & 7.9(8) \times 10^{-13}E^2
   \end{aligned}
   \end{equation}

\noindent and the resulting O--C diagram, Fig. \ref{NSVS_14256825_all_data}, showing a strong peak to peak amplitude reduction with progressing time.  Which is the more representative ephemeris for this binary system remains difficult to determine.  However whilst the five data points from NSVS and ASAS lack precision they do exhibit a predominately decreasing O--C trend and not inconsistent with later data sets.

\section{Conclusions}
We have investigated seven PCEBs listed on one or both the NEA and EPE databases and can find no conclusive evidence to support the presence of circumbinary companions in stable orbits.  The systems selected were not chosen for their interesting ETVs but for being listed on the two databases and for which we could provide additional precise times of minima and so extend the observational timeline.

Two decades have elapsed since ETV methodology was employed to identify circumbinary exoplanets orbiting short period PCEBs and the results so far have been far from encouraging.  With the exception of NY Vir which awaits testing of the new model, \cite{er2021new}, all models predicting circumbinary companions to PCEBs reported here, and other similar systems reported elsewhere, have failed to accurately predict future eclipse times.  It remains possible that some of these systems do have circumbinary companions but their presence is masked by other physical processes e.g. magnetic effects associated with the secondary component.  Magnetic effects have still to be fully understood and are likely to be difficult to predict in the detail necessary to be able to remove them from the observed ETVs, so making the modelling of circumbinary components problematic.
\\
\\
\noindent We find:

   \begin{enumerate}
      \item  Observing ETVs for predicting stable circumbinary companions has proved difficult.  More than 30 models have been compiled over twenty years for seven PCEBs reported here and all have failed with one new model, NY Vir, in need of testing.
      \item AML cannot explain the observed quasi-cyclical ETVs although both apsidal motion and Applegate type mechanisms may, to some extent, contribute to these variations.
      \item Magnetic mechanisms may have significant influence on the ETVs of RR Cae and DE CVn, but none of the other systems. 
      \item Frequently a positive coefficient quadratic term is included in an ephemeris to provide a best fit model.  At present, there are no known processes that can explain this term other than the default explanation of an additional long period circumbinary companion.  Whilst this may be a valid explanation it is likely to require many decades of further observations to be confident of this as a solution.
      \item NSVS and ASAS data sets are often omitted from circumbinary analysis because of their poor precision.  Nonetheless, whilst this data lacks precision it may be considered accurate and so considerably extending the observational timeline.  Omission of this data can lead to very different outcomes in circumbinary modelling.
      \item Increasing observational timelines can lead to more complex circumbinary models with the inclusion of additional circumbinary companions with longer orbital periods. Confidence in a model comes from observing at least two circumbinary orbits which in turn may require further observations and this may now require many decades of observational data.
   \end{enumerate}

\noindent Are eclipse timing variations a reliable indicator of circumbinary companions orbiting short period binary systems?  From a mathematical perspective the ETV methodology cannot be faulted, however when applied to these complex PCEBs the inability to provide accurate long-term solutions leads us to question its reliability.

\section*{Acknowledgements}
This work makes use of observations from the LCO network of telescopes and of the APASS database maintained on the AAVSO website.  We would like to thank Dr Horst Drechsel (Dr. Karl Remeis Observatory), Dr. Roy Ostensen, Andy Baran (Pedagogical University), Tobias Hinse (Korea Astronomy and Space Science Institute), Prof. James Applegate and Dr. Zhu Liying (Yunnan Observatories) who addressed many of the questions we posed. Our thanks go to Prof. Robert Mutel, University of Iowa, for granting access to the Gemini telescope and we would also like to thank the referee whose guidance was greatly appreciated. Lastly, we dedicate this paper to our much missed colleague George Faillace who sadly passed away in 2020. George always wanted us, the Altair Group, to discover an exoplanet using ETVs
%%%%%%%%%%%%%%%%%%%%%%%%%%%%%%%%%%%%%%%%%%%%%%%%%%
\section*{Data Availability}
Eclipse times are published in Appendices \ref{table_A1} and \ref{table_A2} and raw input data is available by contacting lead author.



\bibliographystyle{mnras}
\bibliography{altair} % if your bibtex file is called example.bib

%\begin{thebibliography}{99}
%\end{thebibliography}

%%%%%%%%%%%%%%%%%%%%%%%%%%%%%%%%%%%%%%%%%%%%%%%%%%

%%%%%%%%%%%%%%%%% APPENDICES %%%%%%%%%%%%%%%%%%%%%

\appendix

\onecolumn

\section{Table of Observations}

\begin{center}
\begin{longtable}{cllc|lllc}
\caption{ Compilation of our new times of minima observed between 2018 January and 2022 March. Telescopes used are listed in Table A2.} \\
\toprule
    \multicolumn{1}{p{8.855em}}{BJD\newline{}(+2400000)} & \multicolumn{1}{p{4.43em}}{Cycle} & Filter & Telescope & \multicolumn{1}{p{8.855em}}{BJD\newline{}(+2400000)} & \multicolumn{1}{p{4.43em}}{Cycle} & Filter & Telescope \\
    \midrule
    \textit{DE CVn} & \multicolumn{2}{c}{\textit{T0 = 2452784.55XX}} &       & \multicolumn{1}{c}{\textit{NN Ser}} & \multicolumn{2}{c}{\textit{T0 = 2447344.52XX}} &  \\
    \multicolumn{1}{l}{58867.49839(5)} & 16705 & Clear & 1     & 58928.29271(5) & 89051 & R     & 13\\
    \multicolumn{1}{l}{58877.69444(2)} & 16733 & Clear & 1     & 58971.08908(10) & 89380 & R     & 14 \\
    \multicolumn{1}{l}{58911.55928(2)} & 16826 & Clear & 1     & 59041.85271(1) & 89924 & R     & 14 \\
    \multicolumn{1}{l}{58933.40771(2)} & 16886 & Clear & 1     & 59071.90125(2) & 90155 & R     & 13\\
    \multicolumn{1}{l}{58933.40767(4)} & 16886 & Clear & 2     & 59318.27318(2) & 92049 & Clear & 13\\
    \multicolumn{1}{l}{58965.45193(6)} & 16974 & Clear & 1     & 59358.07772(2) & 92355 & Clear & 13\\
    \multicolumn{1}{l}{59008.78415(14)} & 17093 & IRB   & 6     & 59455.89803(31) & 93107 & IRB   & 8 \\
    \multicolumn{1}{l}{59011.69751(7)} & 17101 & IRB   & 6     & 59625.26242(3) & 94409 & rp    & 15 \\
    \multicolumn{1}{l}{59207.96849(2)} & 17640 & IRB   & 6     & 59625.91290(9) & 94414 & rp    & 16 \\
\cmidrule{5-8}    \multicolumn{1}{l}{59236.73537(11)} & 17719 & Clear & 1     & \multicolumn{1}{c}{\textit{NSVS 14256825}} & \multicolumn{2}{c}{\textit{T0 = 2454274.20XX}} &  \\
    \multicolumn{1}{l}{59280.43222(1)} & 17839 & Clear & 1     & 58742.31827(13) & 40481.5 & Clear & 1 \\
    \multicolumn{1}{l}{59303.37295(3)} & 17902 & Clear & 1     & 58744.36031(2) & 40500 & Clear & 1 \\
    \multicolumn{1}{l}{59319.39511(1)} & 17946 & Clear & 1     & 58759.37118(3) & 40636 & Clear & 2 \\
    \multicolumn{1}{l}{59371.46693(1)} & 18089 & Clear & 1     & 58779.68005(3) & 40820 & IRB   & 6 \\
    \multicolumn{1}{l}{59410.42981(2)} & 18196 & Clear & 1     & 58975.26294(4) & 42592 & IRB   & 10 \\
    \multicolumn{1}{l}{59608.52140(1)} & 18740 & Clear & 3     & 59006.16767(4) & 42872 & IRB   & 10 \\
    \multicolumn{1}{l}{59658.40868(14)} & 18877 & IRB   & 2     & 59025.48311(3) & 43047 & Clear & 2 \\
\cmidrule{1-4}    \textit{HS0705+6700} & \multicolumn{2}{c}{\textit{T0 = 2451822.76XX}} &       & 59061.46504(3) & 43373 & Clear & 1 \\
    \multicolumn{1}{l}{58745.47578(11)} & 72378 & Clear & 2     & 59105.39390(3) & 43771 & Clear & 1 \\
    \multicolumn{1}{l}{58761.64002(6)} & 72547 & Clear & 1     & 59134.31190(2) & 44033 & Clear & 1 \\
    \multicolumn{1}{l}{58777.89999(8)} & 72717 & IRB   & 6     & 59149.32277(1) & 44169 & Clear & 1 \\
    \multicolumn{1}{l}{58784.40406(4)} & 72785 & Clear & 2     & 59173.27394(7) & 44386 & Clear & 1 \\
    \multicolumn{1}{l}{58805.44643(3)} & 73005 & Clear & 2     & 59177.57853(2) & 44425 & IRB   & 6 \\
    \multicolumn{1}{l}{58817.30649(6)} & 73129 & Clear & 1     & 59365.54537(3) & 46128 & Clear & 1 \\
    \multicolumn{1}{l}{58817.40210(4)} & 73130 & Clear & 1     & 59409.47419(3) & 46526 & Clear & 1 \\
    \multicolumn{1}{l}{58817.40219(2)} & 73130 & IRB   & 2     & 59409.47418(2) & 46526 & IRB   & 2 \\
    \multicolumn{1}{l}{58847.33965(5)} & 73443 & Clear & 2     & 59439.49598(2) & 46798 & Clear & 1 \\
    \multicolumn{1}{l}{58861.30401(5)} & 73589 & Clear & 2     & 59464.44049(3) & 47024 & IRB   & 2 \\
    \multicolumn{1}{l}{58893.82398(11)} & 73929 & IRB   & 7     & 59498.32529(3) & 47331 & Clear & 1 \\
    \multicolumn{1}{l}{58911.32735(5)} & 74112 & Clear & 2     & 59541.26079(2) & 47720 & Clear & 1 \\
\cmidrule{5-8}    \multicolumn{1}{l}{58931.41310(7)} & 74322 & Clear & 2     & \multicolumn{1}{c}{\textit{NY Vir}} & \multicolumn{2}{c}{\textit{T0 = 2453174.44XX}} &  \\
    \multicolumn{1}{l}{58952.45547(4)} & 74542 & Clear & 2     & 58142.91303(9) & 49185 & Clear & 4 \\
    \multicolumn{1}{l}{58976.46276(6)} & 74793 & Clear & 2     & 58145.94357(3) & 49215 & Clear & 4 \\
    \multicolumn{1}{l}{59002.47845(11)} & 75065 & Clear & 2     & 58159.98474(4) & 49354 & Sloan r' & 4 \\
    \multicolumn{1}{l}{59068.47492(7)} & 75755 & Clear & 2     & 58190.89567(2) & 49660 & Sloan r' & 4 \\
    \multicolumn{1}{l}{59068.57053(10)} & 75756 & Clear & 2     & 58226.75634(1) & 50015 & Sloan r' & 4 \\
    \multicolumn{1}{l}{59106.54234(8)} & 76153 & Clear & 1     & 58546.87594(4) & 53184 & Sloan r' & 4 \\
    \multicolumn{1}{l}{59132.46253(11)} & 76424 & Clear & 2     & 58580.91825(7) & 53521 & Sloan r' & 4 \\
    \multicolumn{1}{l}{59145.37481(4)} & 76559 & Clear & 2     & 58617.38493(3) & 53882 & IRB   & 12 \\
    \multicolumn{1}{l}{59158.38283(7)} & 76695 & Clear & 2     & 58691.93480(5) & 54620 & V     & 5 \\
    \multicolumn{1}{l}{59178.27745(9)} & 76903 & Clear & 2     & 58692.94499(8) & 54630 & V     & 5 \\
    \multicolumn{1}{l}{59196.25887(6)} & 77091 & Clear & 1     & 58877.19821(9) & 56454 & IRB   & 5 \\
    \multicolumn{1}{l}{59196.35453(3)} & 77092 & Clear & 1     & 58907.09884(4) & 56750 & Clear & 5 \\
    \multicolumn{1}{l}{59196.35456(3)} & 77092 & Clear & 2     & 58907.19987(3) & 56751 & Clear & 5 \\
    \multicolumn{1}{l}{59221.41396(3)} & 77354 & Clear & 2     & 58965.08205(3) & 57324 & IRB   & 5 \\
    \multicolumn{1}{l}{59238.34355(8)} & 77531 & Clear & 2     & 58965.18301(4) & 57325 & IRB   & 5 \\
    \multicolumn{1}{l}{59238.34352(9)} & 77531 & Clear & 1     & 58994.07359(5) & 57611 & IRB   & 5 \\
    \multicolumn{1}{l}{59258.33361(4)} & 77740 & Clear & 1     & 58997.10417(5) & 57641 & IRB   & 5 \\
    \multicolumn{1}{l}{59264.35934(3)} & 77803 & Clear & 1     & 58998.01329(7) & 57650 & IRB   & 5 \\
    \multicolumn{1}{l}{59280.33233(3)} & 77970 & Clear & 1     & 59050.94562(3) & 58174 & IRB   & 5 \\
    \multicolumn{1}{l}{59296.40094(8)} & 78138 & IRB   & 2     & 59203.98484(5) & 59689 & IRB   & 6 \\
    \multicolumn{1}{l}{59318.49543(8)} & 78369 & IRB   & 2     & 59233.17843(3) & 59978 & Clear & 5 \\
    \multicolumn{1}{l}{59364.40578(8)} & 78849 & IRB   & 2     & 59265.90760(7) & 60302 & Sloan r' & 4 \\
    \multicolumn{1}{l}{59451.53985(5)} & 79760 & Clear & 1     & 59266.91788(5) & 60312 & Sloan r' & 4 \\
    \multicolumn{1}{l}{59460.53072(6)} & 79854 & IRB   & 2     & 59289.14130(3) & 60532 & Clear & 5 \\
    \multicolumn{1}{l}{59482.43386(2)} & 80083 & IRB   & 2     & 59296.51549(4) & 60605 & IRB   & 2 \\
    \multicolumn{1}{l}{59494.38970(2)} & 80208 & IRB   & 2     & 59323.18372(3) & 60869 & Clear & 5 \\
    \multicolumn{1}{l}{59509.40616(8)} & 80365 & IRB   & 2     & 59335.40663(5) & 60990 & Clear & 1 \\
    \multicolumn{1}{l}{59541.35223(4)} & 80699 & IRB   & 2     & 59366.41855(4) & 61297 & Clear & 1 \\
    \multicolumn{1}{l}{59570.33321(8)} & 81002 & IRB   & 2     & 59405.91577(3) & 61688 & V     & 2 \\
    \multicolumn{1}{l}{59590.61028(8)} & 81214 & IRB   & 3     & 59434.90738(3) & 61975 & V     & 2 \\
    \multicolumn{1}{l}{59636.42506(4)} & 81693 & IRB   & 2     & 59562.69264(3) & 63240 & IRB   & 3 \\
    \multicolumn{1}{l}{59662.34528(2)} & 81964 & IRB   & 2     & 59596.73499(3) & 63577 & Clear & 1 \\
    \multicolumn{1}{l}{59662.44097(3)} & 81965 & IRB   & 2     & 59612.59452(3) & 63734 & IRB   & 3 \\
    \multicolumn{1}{l}{59670.37961(2)} & 82048 & Clear & 1     & 59612.69552(5) & 63735 & IRB   & 3 \\
\cmidrule{1-4}    \textit{HW Vir} & \multicolumn{2}{c}{\textit{T0 = 2445730.55XX}} &       & 59638.25257(3) & 63988 & IRB   & 11 \\
    \multicolumn{1}{l}{58641.01958(4)} & 110611 & V     & 5     & 59670.47668(3) & 64307 & Clear & 1 \\
\cmidrule{5-8}    \multicolumn{1}{l}{58867.92226(8)} & 112555 & V     & 6     & \multicolumn{1}{c}{\textit{RR Cae}} & \multicolumn{2}{c}{\textit{T0 = 2452723.62XX}} &  \\
    \multicolumn{1}{l}{58908.19048(5)} & 112900 & V     & 8     & 58835.01726(2) & 24076 & IRB   & 10 \\
    \multicolumn{1}{l}{58942.73944(1)} & 113196 & IRB   & 7     & 58862.04709(1) & 24165 & IRB   & 8 \\
    \multicolumn{1}{l}{58983.00767(1)} & 113541 & IRB   & 9     & 58900.01004(7) & 24290 & IRB   & 8 \\
    \multicolumn{1}{l}{59016.97304(2)} & 113832 & V     & 8     & 59116.24708(2) & 25002 & IRB   & 10 \\
    \multicolumn{1}{l}{59049.88795(2)} & 114114 & IRB   & 9     & 59161.19519(2) & 25150 & IRB   & 8 \\
    \multicolumn{1}{l}{59198.00489(4)} & 115383 & IRB   & 6     & 59196.12112(3) & 25265 & IRB   & 8 \\
    \multicolumn{1}{l}{59234.18801(1)} & 115693 & Clear & 5     & 59197.03225(1) & 25268 & IRB   & 8 \\
    \multicolumn{1}{l}{59261.15019(1)} & 115924 & V     & 5     & 59234.08408(3) & 25390 & IRB   & 8 \\
    \multicolumn{1}{l}{59289.16288(1)} & 116164 & V     & 5     & 59265.97296(1) & 25495 & IRB   & 8 \\
    \multicolumn{1}{l}{59325.11249(1)} & 116472 & V     & 5     & 59303.93591(1) & 25620 & IRB   & 8 \\
    \multicolumn{1}{l}{59334.45004(2)} & 116552 & IRB   & 2     & 59423.29148(2) & 26013 & IRB   & 8 \\
    \multicolumn{1}{l}{59379.97069(1)} & 116942 & IRB   & 8     & 59458.21740(4) & 26128 & IRB   & 8 \\
    \multicolumn{1}{l}{59402.96443(2)} & 117139 & V     & 5     & 59496.18038(2) & 26253 & IRB   & 8 \\
    \multicolumn{1}{l}{59561.70293(2)} & 118499 & IRB   & 3     & 59533.23225(3) & 26375 & IRB   & 11 \\
    \multicolumn{1}{l}{59588.66512(3)} & 118730 & IRB   & 3     & 59585.16553(1) & 26546 & IRB   & 11 \\
    \multicolumn{1}{l}{59638.15419(1)} & 119154 & V     & 11    & 59638.92113(7) & 26723 & IRB   & 11 \\
    \bottomrule
  \label{table_A1}%
\end{longtable}%
\end{center}

\setlength{\LTleft}{12pt}

\begin{longtable}{cl}
  \caption{Telescopes used in our observations}\\
    \toprule
    \textbf{Ident} & \multicolumn{1}{c}{\textbf{Telescope}} \\
    \midrule
          &  \\
    1     & 235mm, Ham Observatory West Sussex, UK \\
    2     & 203mm, Woodlands Observatory West Sussex, UK \\
    3     & 104mm OG, PixelSkies, Castillejar, Spain \\
    4     & 508mm Gemini University of Iowa \\
    5     & 508mm Warrumbungle Observatory, Siding Spring \\
    6     & 432mm T21 iTelescope, Mayhill \\
    7     & 508mm T11 iTelescope, Mayhill \\
    8     & 508mm T30 iTelescope, Siding Spring \\
    9     & 432mm T32 iTelescopes Siding Spring \\
    10    & 432mm T17 iTelescope, Siding Spring \\
    11    & 432mm Dubbo Observatory, NSW, Aus \\
    12    & 305mm T18 iTelescope, Nerpio \\
    13    & 2m Faulkes, Siding Spring \\
    14    & 2m Faulkes, Hawaii \\
    15    & 1m Faulkes, Siding Spring \\
    16    & 1m Faulkes, McDonald Observatory \\
    \bottomrule
  \label{table_A2}%
\end{longtable}%

\twocolumn

%%%%%%%%%%%%%%%%%%%%%%%%%%%%%%%%%%%%%%%%%%%%%%%%%%

% Don't change these lines
\bsp	% typesetting comment
\label{lastpage}
\end{document}